# Study of Cherenkov Light Lateral Distribution Function around the Knee Region in Extensive Air Showers

A. A. Al-Rubaiee[1,2,*], U. Hashim[2,*], Marwah M.[1], Y. Al-Douri[2]

[1] *College of Science, Dept. of Physics, the University of Al-Mustansiriyah, 10052 Baghdad, Iraq*
[2] *Institute of Nano Electronic Engineering, University of Malaysia Perlis, 01000 Kangar, Malaysia*

**Abstract**

The Cherenkov light lateral distribution function (LDF) was simulated with the CORSIKA code, in the energy range ($10^{13}$-$10^{16}$) eV. This simulation was performed for conditions and configurations of the Tunka EAS Cherenkov array for two primary particles (p and Fe). Basing on the simulated results, many approximated functions are structured for two primary particles and different zenith angles. This allowed us to reconstruct the EAS events, which is, to determine the type and energy of the primary particles that produced showers from signal amplitudes of Cherenkov radiation which measured with Tunka Cherenkov array experiment. Comparison of the calculated LDF of Cherenkov radiation with that measured at the Tunka EAS array shows the ability for identifying of the primary particle that initiated the EAS cascades determining of its primary energy around the knee region of the cosmic ray spectrum.

## 1. Introduction

The accurate appreciation of the primary cosmic rays (PCRs) spectrum and mass composition in the range around the knee region is very important towards obtaining information about their origin and acceleration mechanisms [1, 2]. In the region of high and ultrahigh energies the only possible way of cosmic rays (CR) registration is indirect from extended air showers (EAS) produced in atmosphere, precisely by registration of atmospheric Cherenkov radiation. The investigation of CRs based on registration of Cherenkov radiation of secondary particles produced in the cascade processes of EAS has intensively been developed in the last years [3, 4]. The Monte Carlo method is one of the necessary tools of numerical simulation for investigation of EAS characteristics and experimental data processing and analysis (determination of the primary particle energy and type and direction of shower axis from the characteristics of Cherenkov radiation of secondary charged particles). Reconstruction of the characteristics of primary particles that initiate the atmospheric cascade from Cherenkov radiation of secondary particles calls for the creation of a library of shower patterns that requires substantial computation time [5].

[*]) For correspondence: Tel: + 6049775021, Fax: + 6049798578, Email: uda@unimap.edu.my, dr.ahmedrubaiee@gmail.com.

The CORSIKA software package [5, 6] is one of the numerical methods that simulate the Cherenkov light LDF emitted by atmospheric cascades initiated by primary high-energy CR protons and nuclei. This simulation requires a long computation time for a single shower with energies $\geq 10^{17}$ eV for a processor with a frequency of a few GHz. Therefore, the development of fast modeling algorithms and the search for approximations of the results of numerical modeling are important practical problems.

Cherenkov light LDF as a function of the distance from the shower axis in EAS was proposed [7, 8]. This function was developed by approximating the results of numerical simulation of LDF of Cherenkov photons emitted by EAS initiated in the Earth's atmosphere by the CR particle as a function of the primary energy and distance from the shower axis [9, 10]. Nerling et al. [11] have used the shower simulation code CORSIKA to investigate the electron energy and angular distributions in high-energy showers. Based on the universality of both distributions, they have developed an analytical description of Cherenkov light emission in extensive air showers, which provides the total number and angular distribution of photons. The parameterization can be used to calculate the contribution of direct and scattered Cherenkov light to shower profiles measured with the air fluorescence technique. On the other side, Leif and Christopher [12, 13] have investigated and parameterized the amount and angular distribution of Cherenkov photons, which are generated by low-energy secondary particles (typically ≲500 MeV), which accompany a muon track in water or ice. These secondary particles originate from small energy loss processes. The authors have elaborated the contributions of the different energy loss processes as a function of the muon energy and the maximum transferred energy. For the calculation of the angular distribution they have developed a generic transformation method, which allows us to derive the angular distribution of Cherenkov photons for an arbitrary distribution of track directions and their velocities. Also, they have followed the same procedure of parameterization by electro-magnetic cascades in water or ice. They have simulated electromagnetic cascades with Geant4 for primary electrons, positrons and photons with energies ranging from 1 GeV to 10 TeV. Additionally, the total Cherenkov-light yield as a function of energy, the longitudinal evolution of the Cherenkov emission along the cascade-axis and the angular distribution of photons are parameterized. Furthermore, they have investigated the fluctuations of the total light yield, the fluctuations in azimuth and changes of the emission with increasing age of the cascade.

In the present work, we have used this parameterization to describe the results of numerical simulation of EAS by the CORSIKA code of Cherenkov light emitted by EAS measured with the Tunka EAS facility, which is constructed to study the energy spectrum and the mass composition of CRs around the knee region [14, 15]. Main advantage of this approach is to reconstruct the real events of Cherenkov radiation measured with Tunka EAS array [16]. Comparison of the approximated Cherenkov light LDF with the reconstructed EAS events registered with the Tunka EAS Cherenkov array allows for primary particle identification and definition of its energy around the knee region.

## 2. Cherenkov Light Production in EAS

The phenomenon of Cherenkov light can be observed when charged particles pass through dielectric medium, such as air, faster than the phase of light in the medium $(v > c/n)$, where $v$ is the speed of the charged particles, $c$ is the speed of light and $n$ is the refractive index.

The threshold energy of charged particles, which excite the Cherenkov radiation through the atmosphere can be determined through the condition $\gamma = \gamma_{th}$ when $\gamma = E/mc^2$, where $\gamma$ is a Lorenz factor of particles in the laboratory system. Lorenz factor at the threshold energy ($\gamma_{th}$) can be defined as:

$$\gamma_{th} = \frac{1}{\sqrt{1-\beta^2}} = \frac{1}{\sqrt{1-(1/n(h))^2}} \qquad (1)$$

The condition $\gamma > \gamma_{th}$ is equivalent to the condition $v > c/n$, i.e. $n\beta > 1$, where $\beta = v/c$, which means that the speed of charged particle should exceed the speed of light $c$ in the medium with refractive index $n$ that depends on the shower height $h$:

$$n(h) = 1 + \zeta(h),$$
$$\zeta(h) = \zeta_0 \exp(-h/h_o), \qquad (2)$$

where $\zeta_0 \approx 3.10^{-4}$, $h_o = 7.5\ Km$. Electron energy at the atmosphere is:

$$E_{th} = mc^2 \gamma_{th} \qquad (3)$$

Substituting Eq.(1) in Eq. (3) yields:

$$E_{th} = \frac{mc^2}{\sqrt{1-(1/n(h))^2}} = mc^2 \frac{n(h)}{\sqrt{n(h)^2-1}} \qquad (4)$$

Equations (2) and (3) give the dependence of the threshold electron energy on the height ($h$) at the atmosphere when:

$$E_{th}(h) = mc^2 \frac{1+\zeta}{\sqrt{2\zeta(1+\zeta/2)}} \approx$$

$$\approx \frac{mc^2}{\sqrt{2\zeta(h)}} = \frac{mc^2}{\sqrt{2\zeta_0}} \exp(h/2h_o) \qquad (5)$$

At the sea level $n = 1 + \zeta_o$, and the Cherenkov radiation will be emitted by electrons that exceed the threshold energy $E_{th}$ i.e. when $\gamma > \gamma_{th} = \frac{E_{th}}{mc^2} \approx 40.8$.

The threshold energy $E_{th}$ for radiation of Cherenkov photons by electrons at the height $(h)$ in the atmosphere can be approximated by Eq. (5) at the height $h_o = 7.5\ Km$ (which means the development of shower at maximum $X_{max} \approx 500 \ g/cm^2$, $E_{th} \approx 34.4 \ MeV$). Cherenkov light produced by very high energy particles ($\beta \sim 1$) slants under a small angle $\theta_r$;

$$\cos \theta_r = \frac{1}{\beta n} \sim 1, \tag{6}$$

where $\theta_r$ is the angle which we can located Cherenkov radiation [17].

Since $\beta \leq 1$, then $\cos \theta_r \leq 1$ and the relation holds when $n > 1$. From Eqs. (1)-(6) we can find:

$$\sin^2 \theta_r = 1 - \frac{1}{(n\beta)^2} \approx 2\zeta(h)\left(\frac{E_{th}^2(h)}{E^2}\right) =$$

$$= 2\zeta_0 \exp(-h/h_o)\left[1 - \frac{E_{th}^2(h)}{E^2}\right], \tag{7}$$

and for $E \gg E_{th}(h)$ at the sea level, we get:

$$\sin \theta \approx \theta_r \approx 1/\gamma_{th} \approx$$

$$\approx \sqrt{2\zeta_0} \exp(-h/2h_o) = 2.45 \cdot 10^{-2} \tag{8}$$

The number of Cherenkov photons per wavelength interval $(\lambda_1, \lambda_2)$ may be obtained from the Tamm-Cherenkov relation under the assumption that the refractive index is independent of the wavelength [11]:

$$\frac{dN_\gamma}{dx} = 2\pi\alpha \sin^2 \theta_r \int_{\lambda_1}^{\lambda_2} \frac{d\lambda}{\lambda^2} =$$

$$= 2\pi\alpha \left(\frac{1}{\lambda_1} - \frac{1}{\lambda_2}\right) \zeta_o \left(1 - \frac{E_{th}^2(h)}{E^2}\right) \exp(-h/h_o), \tag{9}$$

where $\alpha = 1/137$ and

$$\frac{dN_\gamma}{dt} = \frac{dN_\gamma}{dx}\frac{dx}{dt} = x_o \exp(h/h_o)\frac{dN_\gamma}{dx}, \tag{10}$$

where

$$\frac{dx}{dt} = x_o \exp(h/h_o), \tag{11}$$

where $x_o = t_o/\rho_o \approx 3.09 \cdot 10^4$, $x_0$ is the distance of electrons at sea level, $t_o \approx 37.1$ and $\rho_o = 1.2 \cdot 10^{-3}$.

By neglecting the absorption of Cherenkov radiation in the atmosphere, total number of Cherenkov photons $N_\gamma$ radiated by electrons can be written as:

$$N_\gamma = 45 \cdot 10^{10} \frac{E_o}{10^{15} ev} \quad (12)$$

Estimation of the core position and age parameter are also made by using the total number of Cherenkov photons in EAS, which is directly proportional to primary energy $(E_0)$ [16]:

$$N_\gamma = 3.7 \cdot 10^3 \frac{E_o}{\beta_t}, \quad (13)$$

where $\beta_t$ is the critical energy at which it equals ionization losses at the t-unit: $\beta_t = \beta_{ion} t_o$. For electron, $\beta_{ion} = 2.2 \text{ Mev} \cdot (\text{g} \cdot \text{cm}^{-2})^{-1}$, $t_o = 37 \text{ g} \cdot \text{cm}^{-2}$ and $\beta_t = 81.4 \text{ MeV}$ [17].

Thus, the number of Cherenkov photons in the shower is directly proportional to the primary particle energy. It is difficult to measure this value experimentally; therefore, the Cherenkov radiation density, namely, the number of photons per unit area of the detector not model related [18] to the primary particle energy:

$$Q_{(E,R)} = \frac{\Delta N_\gamma (E,R)}{\Delta S}, \quad (14)$$

is used for experimental data processing. As demonstrated by direct measurements of Cherenkov light [17], fluctuations in the form of the LDF for EAS are much less than fluctuations in the photon number $N_\gamma$.

### 3. Results and Discussion
#### 3.1 *Simulation and parameterization of Cherenkov light LDF*

The simulation of Cherenkov light LDF from EAS was performed using the CORSIKA (COsmic Ray SImulations for KAscade) software package [5] with using two models: QGSJET (Quark Gluon String model with JETs) codes [19] were used to model interactions of hadrons with energies exceeding 80 GeV and GHEISHA (Gamma Hadron Electron Interaction SHower) codes [20] were used for lower energies. The CORSIKA code is the Monte Carlo program for simulating EAS for calculations of output hadrons, muons, electrons, and photons in the cascade. The program allows information on the type and energy of shower particles as well as on the angle and the time of their arrival at the observation level to be obtained. The code supports the option of Cherenkov radiation generation by charged shower particles with the use of Electron Gamma Shower (EGS4) system of simulating the electromagnetic component of EAS. The results of simulation by the CORSIKA software package are influenced by such parameters as the primary energy interval, distances passed by Cherenkov radiation, threshold energies of hadrons, electrons, muons and photons; and the factor taking into account multiple Coulomb scattering of electrons. Cherenkov light was simulated for the following values of input parameters of the Tunka EAS facility. The LDF of Cherenkov light was calculated for distances 2.5–400 m from the axis of showers initiated by protons and iron nuclei with energies $10^{13}$–$10^{16}$ eV for different zenith angles of 0°, 10° and 20°.

In order to parameterize the simulated Cherenkov light LDF, the function of four parameters $a$, $b$, $\sigma$, and $r_0$ suggested in [7] was used. This function was normalized by introducing the coefficient $C = 10^3$ m$^{-1}$ providing the correct dimension of $Q$ and consistent with the dimensions of the parameters from [9]:

$$Q(E_0, R) = \frac{Cs \exp[a - \beta]}{b\left[(R/b)^2 + (R - r_o)^2/b^2 + R\sigma^2/b\right]} \text{ m}^{-2}, \quad (15a)$$

where $\beta$ is defined as:

$$\beta = R/b + (R - r_o)/b + (R/b)^2 + (R - r_o)^2/b^2 \quad (15b)$$

where $R$ is the distance from the shower axis and $E$ is the primary particle energy. The values of the parameters $a$, $b$, $\sigma$, and $r_0$ were determined by fitting function (15) to the values of the LDF calculated by the CORSIKA software package on a certain grid ($R$, $E$). The energy dependence of the parameters $a$, $b$, $\sigma$, and $r_0$ were approximated by:

$$k(E) = c_0 + c_1 \log(E/1 \text{ eV}) + c_2 \log(E/1 \text{ eV}) + c_3 \log(E/1 \text{ eV}), \quad (16)$$

here $k(E) = a$, $\log b$, $\log \sigma$, $\log r_0$; $c_0$, $c_1$, $c_2$, and $c_3$ are coefficients obtained using the procedure of approximation for LDF parameters depending on the type of the primary particles (p, Fe) and zenith angle as shown in Fig's 1 and 2.

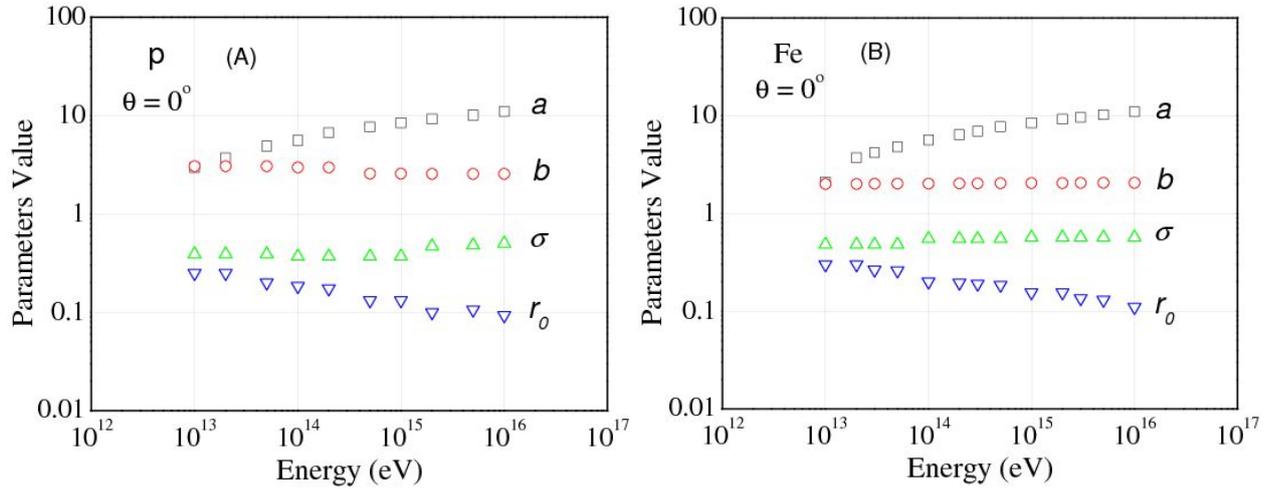

**Fig. 1** Parameter values as a function of the primary energy (Eq. 16) for vertical showers initiated by: (A) primary proton, (B) iron nuclei in the energy range $10^{13}$-$10^{16}$ eV.

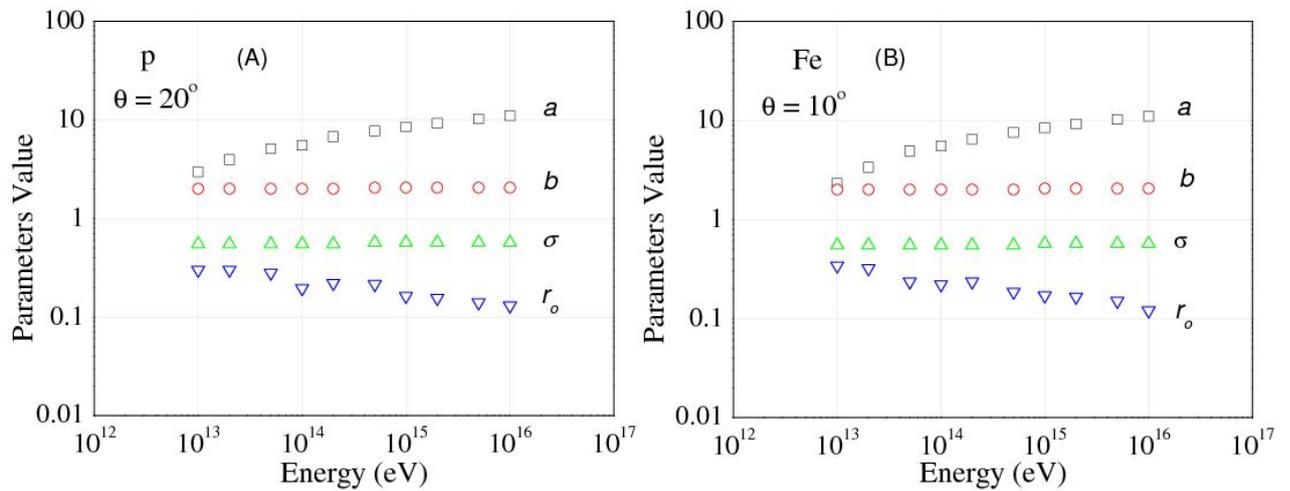

**Fig. 2** Parameter values as a function of the primary energy (Eq. 16) for inclined showers initiated by: (A) primary proton at $\theta = 20°$, (B) iron nuclei at $\theta=10°$ in the energy range $10^{13}$-$10^{16}$ eV.

## 3.2 Reconstruction of showers registered with Tunka EAS array

The total flux of photons from the charged particles of the EAS is proportional to the primary particle energy $E_0$. The amplitude of the signal from the photodetector is proportional to the area of the photocathode and the flux density of the Cherenkov light. The function of the LDF of the Cherenkov radiation describes the dependence of the Cherenkov light on the distance from the shower axis and can be expressed as [16]:

$$N_i = (A_i - A_{0i}) K_a K, \qquad (17)$$

where $N_i$ is the number of photons at $i$-th detector, $A_i$ (channel) is amplitude, $A_{0i}$ (channel) is pedestal base and a range of additional charge-to-digital converter, $K_a$ and $K$ are calibration parameters of photodetectors (in $ns/ch$ units). The time information for event reconstruction of the Tunka EAS array was measured through the following expression:

$$T_i = t_i K_i, \qquad (18)$$

where $t_i$ is the front delay of the maximum amplitude (in $ns$ unit).

Recovered events determine the direction of arrival EAS (zenith $\theta$ and azimuth angle $\phi$), position of the axis in the plane of the detectors (coordinates X and Y):

$$X_s = \frac{\sum_{i=1}^{n_D} N_i X_i}{\sum_{i=1}^{n_D} N_i}, \quad Y_s = \frac{\sum_{i=1}^{n_D} N_i Y_i}{\sum_{i=1}^{n_D} N_i}. \qquad (19)$$

Here $n_D$ is number of detectors, $X_s, Y_s$ (m) are shower coordinates; $X_i, Y_i$ (m) are detector coordinates. The accuracy of core position by using Eq. (19) is about 10 m. The distance to shower axis is defined by the formula:

$$R_i = \sqrt{(X_i - X_s)^2 + (Y_i - Y_s)^2}. \qquad (20)$$

The density of Cherenkov radiation is defined as the number of photons per unit area of the detector ($S_D$) and is given by:

$$Q_i(E, R_i) = \sum_{i=1}^{n_D} N_i(E, R_i)/S_D. \qquad (21)$$

We shall now illustrate the potential of the approach presented here for the reconstruction of events based on time and amplitude characteristics of the signals recorded on the array Tunka [14-16, 21]. The angle of the shower axis (in the approximation of a plane front of the Cherenkov radiation) is recovered by the time characteristics of the signal from the photodetector. Identification of the event was made on the basis of minimizing the function:

$$\Delta = \sum_i \left[ Q^{calc}(E_0, R_i) / Q^{\exp}(R_i) - 1 \right]^2 \to \min . \qquad (22)$$

Trial results (excluding corrections for absorption of Cherenkov light and error estimations for determining the shower axis) of events reconstruction are shown in Fig's 3 and 4.

Figure 3A demonstrates the comparison of calculated Cherenkov light LDF with that measured with Tunka EAS array at the distance 2.5 to 400 m from the shower core. The solid line shows the calculated LDF in vertical showers of primary protons at the energy $8 \cdot 10^{14}$ eV: the hypothesis of p-shower yields $\Delta_{\min} = 0.7710$, while the hypothesis of p-shower with the same energy but for inclined showers $\theta = 10°$ (dashed curve) gives $\Delta_{\min} = 0.8085$.

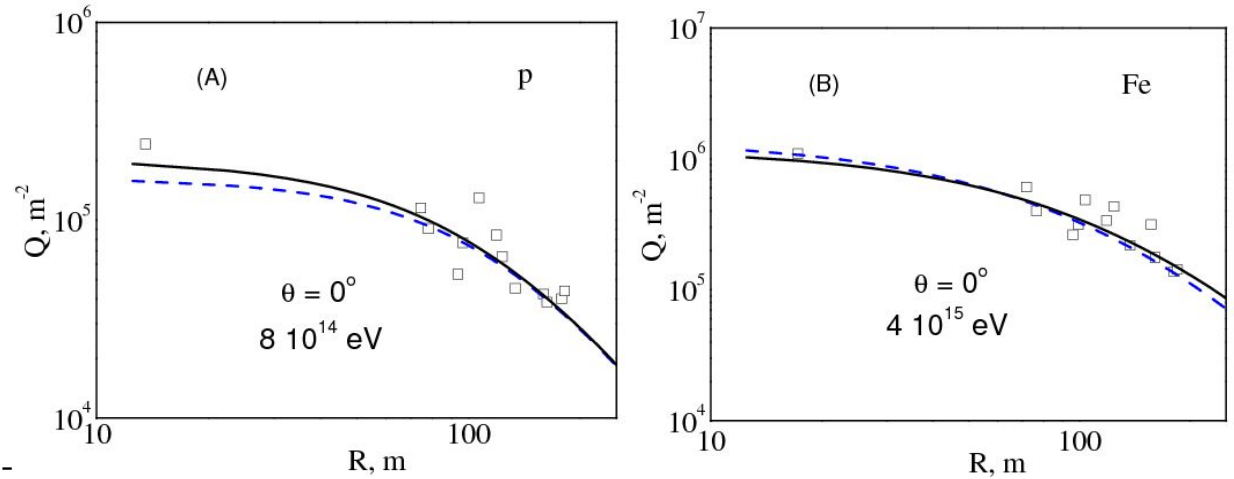

**Fig. 3.** Reconstruction of the vertical showers initiated by: (A) primary proton at the energy $8 \cdot 10^{14}$ eV (solid line) and (B) iron nuclei with energy $4 \cdot 10^{15}$ eV (dashed line), in comparison with Tunka EAS array measurement (symbols).

On the other hand at $\Delta_{min} = 0.6784$ for iron nuclei with energy $4 \cdot 10^{15}$ eV and $\Delta_{min} = 0.7494$ for primary proton with energy $E_0 = 3 \cdot 10^{15}$ eV as seen in Fig. 3B, what means that this event (shower axis inclined at an angle $20^o$ to the vertical) can be identified at about the same level of significance.

The solid line in Fig. 4A shows the calculated Cherenkov light LDF for inclined showers $\theta = 20^o$ of the primary proton with energy 1 PeV, while symbols represent the Cherenkov light LDF as reconstructed from measurements of Tunka array. From the Fig. 4A one can see that p-shower hypothesis gives $\Delta_{min} = 0.7828$ and Fe-shower gives $\Delta_{min} = 0.9465$ at the primary energy 1 PeV (dashed curve), while in Fig. 4B, for inclined showers with $\theta=10^o$ initiated by iron nuclei $\Delta_{min} = 0.7754$ at the energy 5 PeV (solid curve) and $\Delta_{min} = 0.7780$ for primary proton with an energy of 1 PeV (dashed line).

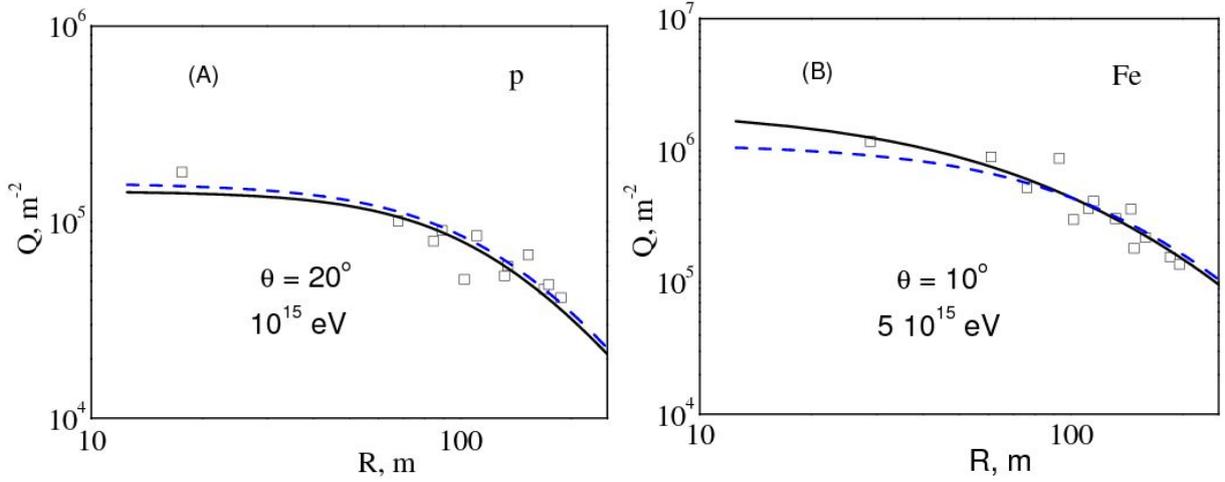

**Fig. 4.** Reconstruction of the inclined showers initiated by: (A) primary proton at the energy $10^{15}$ eV for $\theta=20^o$ (solid line) and (B) iron nuclei with energy $5 \cdot 10^{15}$ eV for $\theta=10^o$ (dashed line), in comparison with Tunka EAS array measurements (symbols).

The accuracy of Cherenkov light LDF in EAS initiated by a primary proton for $0^o$ zenith angle was to be about 10-20% at the distance interval 80-180 m from the shower core and about 5-20% for $10^o$ zenith angle at the same distance interval from the shower core. For iron nuclei the accuracy was found to be 5-15% for $0^o$ zenith angle in the distance interval 80-180 m and 5-20% for $10^o$ zenith angle at the same distance interval.

## 4. Conclusion

The simulation of LDF of Cherenkov radiation in EAS that induced by two primary particles (protons and iron nuclei) was fulfilled using CORSIKA code for configurations and conditions of the Tunka EAS Cherenkov array in the energy range of $10^{13}$-$10^{16}$ eV. On the basis of this simulation, the approximation of the Cherenkov light LDF was developed, which allowed to carry out a trial reconstruction of events. The main feature of the given model consists of the feasibility to make a library of Cherenkov light lateral distribution samples that may be used for analyzing of real EAS events detected with the Tunka EAS array and reconstructing the energy spectrum and chemical composition of the primary cosmic radiations. The comparison of the calculated LDF of the Cherenkov radiation with the experimental data of the EAS Cherenkov Tunka array has demonstrated the potential for primary particle identification and definition of its energy around the knee region.

**Acknowledgments**

Authors thank College of Physics and Tunka Cherenkov EAS experiment collaborators of Irkutsk state University in Russian federation for their support in this work.